\documentclass[11pt]{article}%
\usepackage{amsfonts}
\usepackage{amsmath}%
\usepackage{amssymb}%
\usepackage{graphicx}
\providecommand{\U}[1]{\protect\rule{.1in}{.1in}}

\begin{document}

\title{Quantum non-demolition measurements: Concepts, theory and practice }
\author{Unnikrishnan. C. S.\\\textit{Tata Institute of Fundamental Research, }\\\textit{Homi Bhabha Road, Mumbai 400005, India}}
\date{}
\maketitle

\begin{abstract}
This is a limited overview of quantum non-demolition (QND) measurements, with
brief discussions of illustrative examples meant to clarify the essential
features. In a QND measurement, the predictability of a subsequent value of a
precisely measured observable is maintained and any random back-action from
uncertainty introduced into a noncommuting observable is avoided. The
fundamental ideas, relevant theory and the conditions and scope for
applicability are discussed with some examples. Precision measurements have
indeed gained from developing QND measurements. Some implementations in
quantum optics, gravitational wave detectors and spin-magnetometry are discussed.

\medskip

\noindent Heisenberg Uncertainty, Standard quantum limit, Quantum
non-demolition, Back-action evasion, Squeezing, Gravitational Waves.

\end{abstract}

\section{Introduction}

Precision measurements on physical systems are limited by various sources of
noise. Of these, limits imposed by thermal noise and quantum noise are
fundamental and unavoidable. There are metrological methods developed to
circumvent these limitations in specific situations of measurement. Though the
thermal noise can be reduced by cryogenic techniques and some band-limiting
strategies, quantum noise dictated by the uncertainty relations is universal
and cannot be reduced. However, since it applies to the product of the
uncertainties in non-commuting observables, there is no fundamental limit on
the measurement of one of these observables at the cost of increased
uncertainty and unpredictability in the other. Quantum Non-Demolition
Measurements (QNDM) are those in which repeated measurements of the value of
an observable $O_{1}$ is not hampered by quantum uncertainty generated in any
other physical variable $O_{2}$ as a result of a precision measurement of
$O_{1}$ \cite{Braginsky,Caves,New-review} One may say that a QNDM is achieved
if repeated measurements of $O_{1}$ is possible with predictable results and
if the back action of the uncertainty in $O_{2}$ generated by a measurement of
$O_{1},$ due to the quantum mechanical non-commutativity of the two operators
corresponding to the two observables, is evaded in subsequent measurements of
$O_{1}.$ This class of measurements are also called Back-Action\ Evading (BAE)
measurements. According to an early definition by C. M.\ Caves \cite{Caves},
quantum non-demolition refers to techniques of monitoring a weak force acting
on a harmonic oscillator, the force being so weak that it changes the
oscillator's amplitude by an amount less than the amplitude of the zero-point
fluctuations. A clearer understanding of the basic concept is immediately
achieved if we examine examples cited by V. B. Braginsky \cite{Braginsky},
especially the case of a free particle.

Consider a measurement of the position $x$ of a particle of mass $m,$ with a
precision $\Delta x_{1}.$ Quantum theory does not restrict this precision.
However, such a measurement will introduce an uncontrolled uncertainty of
$\Delta p\geq\hbar/\Delta x_{1}$ in the momentum of the particle. After a
duration $\tau$ the position of the particle is uncertain by $\Delta
x_{2}\simeq\Delta x_{1}+\tau\Delta p/m,$ which could be much larger than
$\Delta x_{1}.$ Hence there is significant back-action on the measurement of
the position. Predictability of the position is demolished because of the back
action of the measurement through the momentum uncertainty. In contrast, the
situation is very different for the measurement of the momentum observable, in
principle. Measurement of momentum $p$ with uncertainty $\Delta p$ does
introduce uncertainty $\Delta x\geq\hbar/\Delta p$ in the subsequent position
of the particle, but this does not feed into the uncertainty of momentum.
$\Delta p_{2}=\Delta p_{1}$, as expected from a conserved constant of motion.

This example serves to define what a QND observable is. If the Hamiltonian of
the system is denoted as $\hat{H}_{s},$free evolution of the system
observables $\hat{O}_{i}$ are given by
\begin{equation}
i\hbar\frac{d\hat{O}_{i}}{dt}=\left[  \hat{O}_{i},\hat{H}_{s}\right]
\end{equation}
To ensure that the uncertainty in $\hat{O}_{i}$ is protected in spite of the
fact that the uncertainty in a conjugate (noncommuting) observable $\hat
{O}_{j}$ will be increased by a measurement of $\hat{O}_{i},$ we need $\left[
\hat{O}_{i},\hat{H}_{s}\right]  =0$ and this implies that $\hat{H}_{s}$ should
not contain an observable $\hat{O}_{j}$ that does not commute with $\hat
{O}_{i}.$ For $\hat{H}_{s}=\hat{p}^{2}/2m,$ the position $\hat{x}$ is not a
QND observable, whereas $\hat{p}$ is.

I stress the caveat that we are still discussing the issue in principle. In
practice, the measurement of the momentum may boil down to the measurement of
position against time (trajectory) and will suffer from back-action. \ One
other point to emphasize is the fact that these measurements do collapse the
wave-function in the usual sense of the phrase, with precision $\Delta x,$
$\Delta p$ etc. and $\Delta x\Delta p\geq\hbar/2.$ Therefore QNDM are not
collapse-evading measurements. Nor they are now-popular weak measurements.

Another instructive example is that of an oscillator, which is archetypical
for serval kinds of real measurements. A quantum mechanical oscillator is
governed by the Hamiltonian
\begin{equation}
H=\frac{\hat{p}^{2}}{2m}+\frac{1}{2}m\omega^{2}\hat{x}^{2}\equiv\left(
a^{\dag}a+\frac{1}{2}\right)  \hbar\omega
\end{equation}
The physical observables obey the uncertainty relation $\Delta x\left(  \Delta
p/m\omega\right)  =\hbar/2m\omega$ with $\Delta x=\Delta p/m\omega$ in a
`coherent state'. This defines the standard quantum limit (SQL),
\begin{equation}
\Delta x=\Delta p/m\omega=\left(  \frac{\hbar}{2m\omega}\right)  ^{1/2}%
\end{equation}
Beating SQL implies squeezing of the uncertainty in one of the variables, at
the expense of the uncertainty in the other.

The oscillator dynamics can be written in terms two corotating conjugate
observables defined by%
\begin{equation}
\hat{x}+i\hat{p}/m\omega=(2\hbar/m\omega)^{1/2}\hat{a}=(X_{1}+iX_{2}%
)\exp(-i\omega t)
\end{equation}
where the complex amplitude $(X_{1}+iX_{2})$ is time independent and hence a
constant of motion.%

\begin{align}
\hat{X}_{1}  &  =\hat{x}\cos\omega t-(\hat{p}/m\omega)\sin\omega t\nonumber\\
\hat{X}_{2}  &  =\hat{x}\sin\omega t+(\hat{p}/m\omega)\cos\omega t
\end{align}
with $\Delta\hat{X}_{1}\Delta\hat{X}_{2}\geq\hbar/2m\omega$.

The crucial difference between the observable pair $\left(  \hat{x},\hat
{p}\right)  $ and $\left(  \hat{X}_{1},\hat{X}_{2}\right)  $, both of which
obey the uncertainty relation, is that while the first pair has back-action
dependence through the equation of motion with the free Hamiltonian $\hat
{H}_{0}$ that depends quadratically on them,
\begin{equation}
\frac{d\hat{x}}{dt}=-\frac{i}{\hbar}\left[  \hat{x},\hat{H}_{0}\right]
\end{equation}
the second pair are both constants of motion,%
\begin{equation}
\frac{d\hat{X}_{i}}{dt}=\frac{\partial\hat{X}_{i}}{\partial t}-\frac{i}{\hbar
}\left[  \hat{X}_{i},\hat{H}_{0}\right]  =0
\end{equation}
($\hat{x}$ and $\hat{p}$ are time dependent whereas $\hat{X}_{1}$ and $\hat
{X}_{2}$ are not). Therefore, if an interaction Hamiltonian $H_{I}$ such that
$\left[  \hat{X}_{1},\hat{H}\right]  =0$ can be designed for the measurement
of $\hat{X}_{1},$ the observable can be measured without back action from
$\hat{X}_{2}$, which of course is disturbed by the measurement of $\hat{X}%
_{1}.$

\section{What QNDM are not!}

It is perhaps important to state briefly what QNDM are not and this seems
necessary in the context of some dismissive views expressed about the
essential idea, possibly generated by the way some measurements tries to
achieve a QNDM. An early discussion about the context and definition is in
reference \cite{Braginsky}, which stressed the aspect of multiple measurements
on the same physical system without introducing measurement induced quantum
uncertainty into the observable being measured. The essence of that discussion
is that a QND measurement aims to identify and measure a metrologically
relevant variable for which deterministic predictability of its possibly time
dependent values are not demolished and obliterated by the quantum uncertainty
introduced into another non-commuting variable as a result of the measurement
of the first variable. \ In particular, the idea is very different in context
from making repeated measurements of the same variable on a microscopic
(atomic) quantum system, as in the measurement of the spin projection of an
electron in a particular direction, which gives the same predictable result
after the first unpredictable measurement. Limitations from quantum mechanics
are to be considered not because the system itself is microscopic and atomic,
but because the physical system, often macroscopic, is near its quantum ground
state or its energy levels relevant for metrology need to be resolved below
the zero-point contribution. The original context is detection of
gravitational waves with resonant bar detectors, where it was necessary to
devise methods to monitor displacement amplitudes less than $10^{-20}$ m of
the end of a macroscopic mass weighing a ton or more, with measurement
bandwidth of 1 kHz ($\tau\simeq10^{-3}$ s) or so. This is comparable to the
quantum zero-point motion of such a metal bar. A measurement with $\Delta
x_{1}\leq10^{-20}$ m introduces uncertainty of $\Delta v\geq\hbar\tau/m\Delta
x_{1}\simeq10^{-20}$ m/s, which will obliterate a reliable second measurement.
\textquotedblleft The first measurement plus the subsequent free motion of the
bar has `demolished' the possibility of making a second measurement of the
same precision...\textquotedblright\ This may be contrasted with a recent
criticism of QNDM \cite{Monroe}, with title \textquotedblleft Demolishing
quantum non-demolition\textquotedblright. \ 

\begin{quotation}
If one already knows that the system is in a particular eigenstate of the
measuring device, then, obviously, a measurement on the system will produce
that eigenstate and leave the system intact. Zero information is gained from
the repeated measurement. On the other hand, when the system is not in an
eigenstate of the measuring device, the quantum state can be thought to
collapse to one of its eigenstates...

In that case, information is gained from the system, and the QND measurement
most certainly demolishes the system. The concept of QND measurement adds
nothing to the usual rules of quantum measurement, regardless of interpretation...

As a common example of an imperfect measurement, consider
photodetection...Sure, the photon has disappeared, but if our detector
indicates that we had one photon, we can always create another and get the
same answer again and again, exactly like a QND measurement... In every case,
the concept of QND measurement is confusing and unnecessary. Why not demolish
the term \textquotedblleft QND\textquotedblright?
\end{quotation}

Why is it that all the serious literature of QNDM so easily dismissed?
Unfortunately, what is referred to in this critical note is not QNDM at all in
any of its forms! Such is the confusion in spite of clear examples is in fact
a surprise for me, personally. However, in the context of this short review,
it suffices to say that QNDM is a distinct and useful idea within the premises
of standard quantum measurement practise and its conceptual strength will be
assessed properly only after one manages to measure quantities that are at
present impossible to measure otherwise. The need to keep the physical state
undemolished in a QNDM is to monitor and measure its tiny changes due to an
external interaction, with precision possibly below the standard quantum
limit. Indeed, the abstract of a seminal paper \cite{Braginsky} reads,
\textquotedblleft some future gravitational-wave antennas will be cylinders of
mass approximately 100 kilograms, whose end-to-end vibrations must be measured
so accurately ($10^{-19}$ centimeter) that they behave quantum mechanically.
Moreover, the vibration amplitude must be measured over and over again without
perturbing it (quantum nondemolition measurement). This contrasts with quantum
chemistry, quantum optics, or atomic, nuclear, and elementary particle
physics, where one usually makes measurements on an ensemble of identical
objects and does not care whether any single object is perturbed or destroyed
by the measurement...\textquotedblright.

The key point is that while the measurement involves quantum mechanical
constraints and limitations, like the uncertainty principle, the single
physical system on which repeated measurements are to be made need not be
microscopic. More importantly, the value of the physical observable is
expected to change during the repeated the measurement and that is precisely
what is being monitored without back action of the quantum uncertainty - there
is no metrological interest in the repeated measurements of a quantity that is
known to remain a constant!

\section{Generalized QNDM}

The basic idea of QNDM can be expanded in a useful way to bring in a larger
class of measurements. All practical implementation of such a generalized
picture of QNDM involves the measurements of a system variable without
significantly affecting the key observable of the system by coupling an
auxiliary variable of a `probe' system to the `signal' such that an observable
of the probe faithfully represents the signal observable (figure 1). The probe
observable is measured by a `meter' or detector by direct interaction such
that quantum disturbance created in the probe variable as a result of the
measurement does not feed back into the signal in spite of the coupling.
Typically this implies that the signal and probe variables are conjugate
pairs, but belonging to two different physical systems (physically both the
signal and the probe may be of the same physical nature, like light). The
conventional `collapse' happens in the interaction of the probe and meter and
not in the interaction of the system and the probe. In some discussions the
term `meter' is used to refer to the probe-meter system together.%

\begin{figure}[ptb]%
\centering
\includegraphics[
height=0.8466in,
width=2.808in
]%
{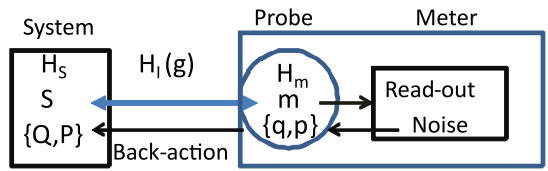}%
\caption{Scheme of a quantum measurement. See text for details. The final
stage of coupling a classical meter to the probe involves collapse of the
state as well as injection of quantum and other noise back into the probe
system. A proper choice of the probe observable avoid back-action on the
signal. }%
\end{figure}

We can now write down the mathematical requirements for the definition of a
QNDM. The requirement that the signal variable represented by the quantum
mechanical operator $\hat{A}(t)$ is deterministically predictable implies
that
\begin{equation}
\left[  \hat{A}(t_{j}),\hat{A}(t_{i})\right]  =0
\end{equation}
for different times $t_{k}.$ For example, for the free particle, momentum
satisfies this relation, being a constant of motion. For an oscillator, the
position and momentum are
\begin{align}
\left[  \hat{x}(t),\hat{x}(t+\tau)\right]   &  =\frac{i\hbar}{m\omega}%
\sin\omega\tau\nonumber\\
\left[  \hat{p}(t),\hat{p}(t+\tau)\right]   &  =i\hbar m\omega\sin\omega\tau
\end{align}
The commutators are zero only at specific instants separated by a half-period,
for each observable, and they\ are called stroboscopic QND variables.
Labelling two noncommuting system observables as $\hat{S}_{i}\equiv\{\hat
{Q},\hat{P}\}$ and the probe-meter observables as $\hat{m}_{j}\equiv\{\hat
{q},\hat{p}\},$ with their own Hamiltonian evolutions and an interaction
Hamiltonian $\hat{H}_{I}$ for the coupling between the system and the meter,
\begin{align}
i\hbar\frac{d\hat{S}_{i}}{dt}  &  =\left[  \hat{S}_{i},\hat{H}_{s}\right]
-\left[  \hat{H}_{I},\hat{S}_{i}\right] \nonumber\\
i\hbar\frac{d\hat{m}_{j}}{dt}  &  =\left[  \hat{m}_{j},\hat{H}_{m}\right]
-\left[  \hat{H}_{I},\hat{m}_{j}\right]
\end{align}
While the observable $\hat{S}_{i}$ could be time dependent, as in the case of
the position of a mirror due to the interaction with a passing gravitational
wave, QNDM demands that it does not change due to the interaction with the
meter system that is used to read out the value of the variable. So, \ a QND
observable of the system satisfies $\left[  \hat{S}_{i},\hat{H}_{s}\right]
=0.$ For the same observable to be back-action evading (BAE), it should
satisfy $\left[  \hat{S}_{i},\hat{H}_{I}\right]  =0.$ Since we want the meter
observable $\hat{m}_{j}$ to change due to the coupling to the system, for an
efficient measurement, $\left[  \hat{H}_{I},\hat{m}_{j}\right]  \neq0.$ Taking
the QND observable $\hat{S}_{i}$ as $\hat{Q},$ these requirements suggest that
the meter observable for readout should be $\hat{p}$ and that the interaction
Hamiltonian could be of the form
\begin{equation}
\hat{H}_{I}=g\hat{Q}\hat{q}%
\end{equation}

The back action from the meter is evaded by choosing the system and meter
observables with a conjugate nature, like intensity of the signal beam and
phase of the meter beam in an optical QNDM. For example, in an optical QNDM,
the system observable could be the intensity and the phase of the probe beam
the readout observable, with an interaction Hamiltonian $\hat{H}_{I}=\chi
\hat{n}_{s}\hat{n}_{p},$ where $\chi$ is the optical Kerr nonlinearity. For
the measurement to qualify as a `good' measurement, the correlation between
the variations in the signal and the probe has to be high enough, ideally
unity. \ This is achieved by choosing the right Hamiltonian and the coupling
$g,$ keeping in mind that the choice is constrained by the need to evade back-action.

\section{Demonstrations}

Several demonstrations of QNDM are now available, mainly in the context of
quantum noise limited measurements in several areas of optics and atomic
physics. There have been some demonstrations that are in tune with the
development of original ideas in QNDM, for macroscopic mechanical systems
observed close to their quantum ground state where quantum noise is readily
observable. We mention a limited sample to clarify the essential concepts.
However, we omit the details of implementation and analysis and refer to the
relevant papers for \ details.

\subsection{Opto-Mechanical System}

In this example, the metrological goal is to monitor the quantum radiation
pressure noise of an optical signal beam by its mechanical effect on the
position of macroscopic mass attached to a spring, forming a classical
oscillator (or a quantum oscillator with extremely small spacing in the
quantized energy). A natural choice for the meter is another weak optical
beam. The coupling between the signal and meter is achieved by the device of
an optical cavity with which both light fields are resonant (figure 2). The
macroscopic mass oscillator is one of the mirrors of the cavity in the QNDM
implementation \cite{Heidman,Walls}. Then the intensity fluctuations of the
signal, either due to a modulation or due to quantum fluctuations (radiation
noise pressure), will translate to displacement noise of the mirror. However,
since the meter field is resonant with the cavity, the intensity of the
reflected field is unaffected to first order in displacement, but the phase of
the meter beam (with weak intensity) is linearly affected. This enables a
faithful measurement of the signal beam intensity variations, without any back
action on the intensity of the signal beam, through the signal obtained by
forming an optical cavity with the oscillator mass as one of the mirrors. The
physical system itself resembles closely the configurations in interferometric
gravitational wave detectors where the actual signal is the displacement $x$
of the mirror that is measured as first order phase changes in the probe
light, but affected by the radiation pressure noise through the interaction
Hamiltonian of the form $H_{I}=g\hat{n}\hat{x}$ where $\hat{n}$ is the photon
number operator. (Interaction Hamiltonian of the form $H_{I}=\lambda\hat
{F}\hat{x}$ is generic for measurement of weak forces).%

\begin{figure}[ptb]%
\centering
\includegraphics[
height=0.6175in,
width=2.476in
]%
{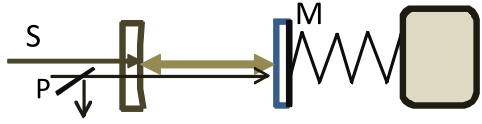}%
\caption{Radiation pressure of the signal beam (S) causes fluctuations in the
position of the mirror on spring (M) and in turn changes the phase of the
resonant weak probe beam (P), in the cavity configuration.}%
\end{figure}

The coupling between the signal and probe beams has been implemented in
several experiments employing the non-linear optical effects inside the cavity.

Successful implementations are considerably more complicated, done at
cryogenic temperature, involving Hamiltonians nonlinear in the observables
\cite{Hertzberg} (in contrast to bilinear Hamiltonians with coupling
coefficients representing a nonlinearity). The most important metrological
context for optomechanical QNDM is the detection of gravitational waves with
advanced optical interferometers, which I will discuss later.

\subsection{Optical QNDM and the quantum tap}

As in other QNDM schemes in practise, optical QNDM couples a meter beam to a
signal beam, typically through an atomic medium and then the strong
correlation between the meter observable and the signal observable is used for
a measurement of the signal by a real measurement on the meter beam
\cite{Grangier-R,Grangier1,Kimble}. The observables are chosen such that there
is no back action. Optical QNDM makes use of nonlinear interaction between a
signal beam and a meter beam, through a generalized Kerr effect -- \ intensity
dependent\ changes in the effective refractive index, $n=n_{0}+n_{2}I,$ due to
the presence of the optical beam with intensity $I$. This is characterized by
a nonlinear phase shift proportional to intensity,
\begin{equation}
\phi_{i}=\frac{2\pi l_{i}}{\lambda_{i}}n_{2i}I_{i}%
\end{equation}
where the index refers to either $s$ or $m,$ signal beam or meter beam. The
cross gain for the coupled system is
\begin{equation}
g=2\sqrt{\phi_{m}\phi_{s}}%
\end{equation}
which defines how the fluctuations in one beam feeds into the other. Denoting
the fluctuations in amplitude and phase quadratures as $\delta X$ and $\delta
Y$, we have%
\begin{align}
\delta X_{o}^{s}  &  =\delta X_{i}^{s}\qquad\delta Y_{o}^{s}=\delta Y_{i}%
^{s}-g\delta X_{i}^{m}\nonumber\\
\delta X_{o}^{m}  &  =\delta X_{i}^{m}\qquad\delta Y_{o}^{m}=\delta Y_{i}%
^{m}-g\delta X_{i}^{s}%
\end{align}
because the two intensities are decoupled, but the phases are coupled. The
amplitude quadrature fluctuation is $\delta X=\delta n/\sqrt{n}$ and the phase
quadrature is $\delta Y=2\delta\phi\sqrt{n},$ where $n$ is the number of photons.

\begin{figure}[ptb]%
\centering
\includegraphics[
height=1.0039in,
width=3.4767in
]%
{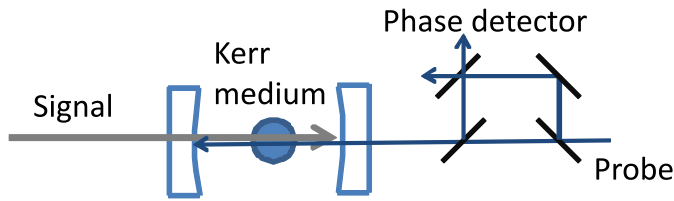}%
\caption{Schematic diagram of an optical QNDM. The strong signal and weak
probe beams interact via a Kerr nonlinearity in the atomic medium, causing a
change in the phase of the probe proportional to the intensity of the signal.
Intensity of the signal beam is not affected by the increased uncertainty in
the amplitude quadrature of the probe due to the precision phase measurement
changes only the phase of the signal and not its amplitude, enabling
back-action evasion. }%
\label{Kerr}%
\end{figure}

Since the intensity variations cause only a change in the phase, and not the
intensity, of the coupled beam, back-action is evaded. The modulations of the
signal beam can be measured as modulations of the phase of the meter beam
without affecting the intensity of the signal beam. Though the intensity noise
in the meter beam does affect the phase of the signal beam, it does not feed
into the other quadrature that is being monitored. Naturally, an
interferometric set up in which the phase of the meter beam is measured with
reference to the stable reference of a split-off part of the meter beam is
required (figure \ref{Kerr}).%

Criteria for an optical QNDM have been developed and discussed in ref.
\cite{Holland,Roch,Grangier-R}. Since quantum noise is unavoidable, one
usually has $\Delta X_{s}\Delta X_{m}\geq1$ with the equality achieved at SQL.
A QND measurement is characterized by $\Delta X_{s}\Delta X_{m}<1.$ Defining
the signal-to-noise ratio as $R=$ $\left\langle X\right\rangle ^{2}%
/\left\langle \delta X\right\rangle ^{2}$ for the various beams, the goal is
to minimize additional noise in the interaction of the signal and meter such
that the transfer function for $R$ from input to output ($T=R_{out}/R_{in}$)
is as close as possible to unity. For an ideal classical beam splitter (a
classical optical tap), for example, with transmissivity $t^{2},$ $X_{out}%
^{s}=t^{2}X_{in}^{s}$ and the meter output will have the rest of the signal
beam, $(1-t^{2})X_{in}^{s}.$ Hence $T_{s}+T_{m}=1$ and no classical device can
exceed this. However, $\Delta X_{s}\Delta X_{m}<1$ implies $T_{s}+T_{m}\geq1$
and ideal QND can approach $T_{s}+T_{m}=2.$ One implementation of these ideas,
with $T_{s}+T_{m}>1$, was realized with the nonlinear coupling generated using
a three level atom in which the ground state is \ coupled to the strong
transition by the detuned weak probe beam and the level 2 to 3 \ in the ladder
by the strong signal beam \cite{Grangier1,Grangier2}. This scheme avoids
absorption from the signal beam, yet preserving the strong coupling between
the signal and the probe, providing a phase shift of the probe proportional to
the intensity of the signal beam. \ Intensity of the signal beam is not
affected by the increased uncertainty in the amplitude quadrature of the probe
due to the precision phase measurement because it changes only the phase of
the signal and not its amplitude, again through the Kerr coupling, enabling
back-action evasion.

\subsection{Atomic spin systems}

Atomic spin systems offer a metrologically important physical scenario for
implementing and testing QND measurement schemes \cite{Japanese,Romalis2}. For
individual atomic spins the projections along different directions are
noncommuting observables. For a spin ensemble, with total spin $S=(S_{i}%
,S_{j},S_{k}$),
\begin{equation}
\left\langle \Delta S_{i}^{2}\right\rangle \left\langle \Delta S_{j}%
^{2}\right\rangle \geq\frac{1}{4}\left\langle \Delta S_{k}^{2}\right\rangle
\end{equation}
A coherent spin state is one that satisfied the minimum uncertainty with equal
uncertainties in the two directions. Therefore, the spin state is considered
squeezed when one of the uncertainties, $\left\langle \Delta S_{i}%
\right\rangle <\frac{1}{2}\left\langle S_{k}\right\rangle .$ This is
consistent with the idea that a spin system polarized along a particular
direction, the spin noise (variance) scales as number of spins $N$. The
elementary spin being $\hbar/2$ with variance $\hbar^{2}/4,$ the spin $S$ is
worth $2N$ elementary spins and hence the variance of uncorrelated spins is
$S/2.$ Squeezing then involves generating correlations among the elementary
spin by an interaction. A measurement of one projection with a precision
$\left\langle \Delta S_{x}\right\rangle <\frac{1}{2}\left\langle
S_{k}\right\rangle $ results in a spin-squeezed state with increased
uncertainties in the other projections (a weaker condition $\left\langle
\Delta S_{x}\right\rangle <\frac{1}{2}\left\langle S\right\rangle $ was shown
to be sufficient for increased bandwidth of measurements at the quantum limit
\cite{Romalis1}). \ The conditions on the Hamiltonian of the system $S$ and
the probe $m$ are obvious,
\begin{align}
\left[  S_{z},H_{S}\right]   &  =0;\quad\text{Ensures }\left[  S_{z}%
(t_{2}),S_{z}(t_{1})\right]  =0\nonumber\\
\left[  S_{z},H_{I}\right]   &  =0;\quad\text{Ensures BAE}\nonumber\\
\left[  s_{z}(m),H_{I}\right]   &  \neq0;\quad\text{Ensures that }%
s_{z}(m)\text{ is a valid probe}%
\end{align}
and this suggests $H_{I}=\alpha S_{z}s_{z}(m).$

Precision magnetometry with sensitivity reaching a femto-Tesla is a motivating
factor for QNDM on spin ensembles. The fundamental noise is the quantum spin
shot noise with SQL variance of $S/2$ for the spin-$S$ ensemble. The basic
measurement scheme involves the Larmour precession of the spins in a weak
magnetic field which can modulate the polarization of a weak linearly
polarized probe beam that is detuned from the hyperfine resonances. With no
net polarization, one obtains a polarimetric signal of the quantum noise at
the Larmour frequency \cite{Romalis1,Crooker}. The goal is to implement a QNDM
of a magnetometer signal, which is the Larmour precession of the coherent
polarization generated in the atomic vapour with a circularly polarized pump
beam. Implementation of \ QND measurement with a stroboscopic BAE scheme in
atomic vapor of Potassium is discussed in reference \cite{Romalis2}.

\subsection{QNDM of photon number in a cavity}

An impressive application of the QND idea that goes beyond demonstration of
principles and strategies is that of the measurement of the number of photons
inside a high finesse optical cavity, without altering this number by
absorption, by observation of the change in the phase of atomic states of a
passing atomic beam that interacts with the photons inside the cavity
\cite{Haroche}. The Stark shift (light shift) induced splitting of the energy
levels of the atom in the cavity containing $n$ photons is (obtainable from
the Jaynes-Cummings model) is
\begin{equation}
\Delta E=\frac{\hbar\Omega^{2}}{2\Delta}n+\frac{\hbar\Omega^{2}}{4\Delta}%
\end{equation}
which results in an n-dependent discrete phase shift,%
\begin{equation}
\Phi_{c}(n)=\frac{\Omega^{2}\tau}{2\Delta}n
\end{equation}%
\begin{figure}[h]%
\centering
\includegraphics[
height=0.8465in,
width=3.4447in
]%
{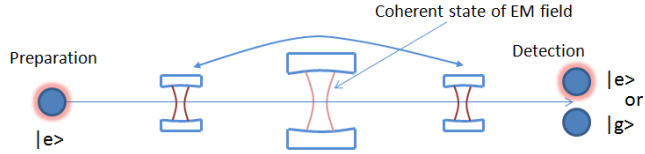}%
\caption{QND-BAE measurement of the number of microwave quanta in the cavity.
The central cavity has a small number of photons that change the relative
phase of the superposition of the excited and ground states of the passing
atoms. The two auxiliary cavities define a Ramsey interferometer with a
scannable relative phase. Final state selective detection enables an iterative
determination of the number state inside the main cavity. See text for more
details. }%
\label{Ramsey}%
\end{figure}
The experiment is implemented as a Ramsey interferometer with three microwave
cavities, with the two auxiliary cavities for state preparation and analysis
with a precisely tunable phase difference between them (figure \ref{Ramsey}).
\ A `Pi/2' pulse of microwave radiation is applied in the first cavity to
atoms prepared in the excited state, which changes the state to a coherent
superposition of the ground and excited states. The state will evolve due to
free evolution as well as due to the phase acquired in the cavity. The final
state of the atoms ($e$ or $g$) is detected after a second $\pi/2$ pulse in
the final cavity with tunable Ramsey phase $\Phi.$ Scanning the Ramsey phase
results in sinusoidal modulation of the average fraction of the two atomic
states and of the probability of detection in a particular state (figure 5).
For example,
\begin{equation}
P_{\left\vert e\right\rangle \rightarrow\left\vert g\right\rangle }=\cos
^{2}\left(  \frac{\Phi_{c}(n)+\Phi}{2}\right)
\end{equation}
%

\begin{figure}[h]%
\centering
\includegraphics[
height=1.1044in,
width=2.0567in
]%
{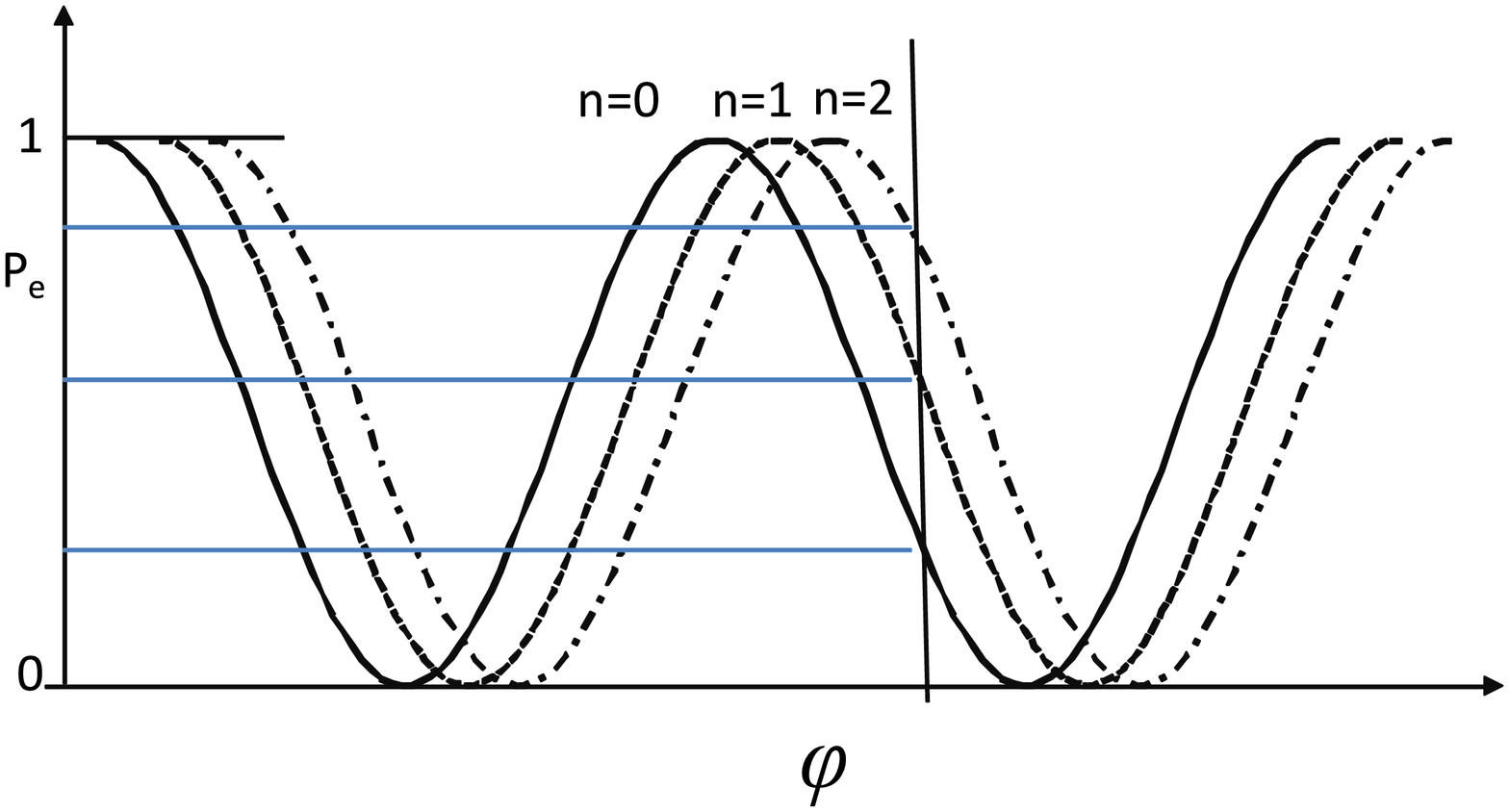}%
\caption{The probability to get a particular final state as a function of the
Ramsey phase. The three curves are for three different photons numbers inside
the cavity.}%
\end{figure}
There are two observables, atom in the ground state and atom in the excited
state, which are complimentary. Since the probability depends on the discrete
number of photons in the main cavity, the sinusoidal probability curve will
shift in phase by a discrete jump when one photon is added or subtracted from
the cavity. Therefore, each set of measurements of $P_{\left\vert
e\right\rangle \rightarrow\left\vert g\right\rangle }$ or $P_{\left\vert
e\right\rangle \rightarrow\left\vert e\right\rangle }$determines the photon
number probabilistically.

If the atom prepared in a excited state comes out in excited state after the
interaction with the cavity, then its phase is shifted by $0$ or $2\pi$ and
the photon number in the cavity is most probably $0$ or $n$, with sinusoidal
variation of the the probability for other photons numbers (the interaction is
tuned to get a particular predetermined phase shift of $2\pi$ for $n$
photons). If the atom is detected in the ground state the phase is $\pi/2$ and
the probability peaks at photons number $n/2.$ Since the detuning is large,
only the phase of the atoms is affected and there is no photon absorption or
stimulated emission, maintaining the QND nature of the measurement. The
interaction with the atoms does feed back to the phase of the cavity field,
but that does have any back action on the photon number.

In this example, the observables do not return definite values, but only a
probability distribution. The measurement is characterized as a two-element
POVM (Positive Operator Valued Measure) $S_{j}$ corresponding the state of the
detected atom ($S_{0}+S_{1}=I)$, which in turn determines a partial
(probabilistic) measurement of the photon number ($\hat{n}=a^{\dag}a$) in the
cavity.
\begin{equation}
S_{j}=\cos^{2}\left(  \frac{\Phi+\Phi(a^{\dag}a)-j\pi}{2}\right)
\end{equation}
If $\rho$ the initial state of the field, the probability of finding the atom
in state $j$ is%
\begin{equation}
P_{j}(\rho_{i})=Tr(\rho S_{j})
\end{equation}
A detection of the atom in state $j$ projects the field state to%
\begin{equation}
\rho_{p}(j)=\frac{\sqrt{S_{j}}\rho\sqrt{S_{j}}}{Tr(\rho S_{j})}%
\end{equation}
One is effectively starting with a uniform initial density matrix (probability
being equal for photon numbers from $0$ to $n$) and then building up $\rho
_{p}(j)$ in repeated QND measurements. This is one case where repeated
measurement without demolition of the state is achieved with new information
gained in each step of the experiment, providing a strong counter example to
the criticism expressed in ref. \cite{Monroe}.

Braginsky, who is one of the originators of QNDM idea, remarked about these
measurements, \cite{Brag-Adolescent}

\begin{quotation}
Several years ago, S. Haroche and his colleagues successfully demonstrated
absorption-free counting of microwave quanta. In my opinion, this is one of
the most outstanding experiments conducted during the second half of the 20th century.
\end{quotation}

\subsection{Squeezed light in gravitational wave detection}

Since the focus has now shifted from resonant metal oscillator detectors to
optical interferometers for the detection of gravitational waves, beating the
standard quantum limit for measurements also is focused in the optical domain,
specifically in the use of quantum noise-squeezed light and its vacuum state.
Indeed this direction of research has turned out to be successful in practical
terms for the GW detector and the advanced interferometer detectors that are
being commissioned for observations have been tested with squeezed light, with
promising benefits in sensitivity and stability of operation. Referring back
to our discussion on QND with a mechanical oscillator and light, we can sketch
the basic idea. The gravitational wave causes small oscillations of the
suspended mirrors of the optical cavity and this causes first order changes in
the phase of the stored light and only second order changes in its intensity
(being locked to the peak of a Fabry-Perot resonance). Hence the gravitational
wave signal is in the phase quadrature, contaminated by the minimum
uncertainty noise in the same quadrature of the coherent state vacuum. The
noise in the intensity quadrature is radiation pressure noise that affects the
position of the mirror, causing additional noise in the phase quadrature, if
large. The detection shot noise, in the phase quadrature relevant for the
interferometer sensitivity, decreases as $\bar{n}^{-1/2},$ where $\bar{n}$ is
the average number of photons in the detection band, whereas radiation
pressure noise on the mirror is the fluctuation in the momentum transfer
($p=2\bar{n}h\nu/c)$ and increases as $\bar{n}^{1/2}.$ The two variances add
and determine the SQL. However, the radiation pressure noise is frequency
dependent when translated into the actual mirror motion because the mirrors
are suspended as pendula and the response decreases as $1/f^{2}$ where $f$ is
the natural frequency.%

\begin{figure}[ptb]%
\centering
\includegraphics[
height=1.4627in,
width=2.4969in
]%
{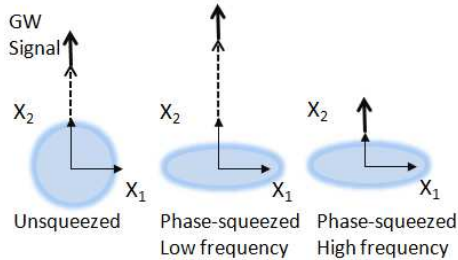}%
\caption{Noise reduction by squeezing the vacuum noise, shown here for
squeezing in the phase quadrature. The GW signal is in the phase quadrature
(X$_{2}$) and its measurement is limited by the quantum shot noise as well as
the radiation pressure noise (dotted arrow). Squeezing the phase quadrature
reduces phase noise and improves the sensitivity to GW, but it also increases
the radiation pressure noise because the amplitude (X$_{1}$) uncertainty
increases (back-action). This extra noise is negigible at high frequency
because of the mirror pendulum response, though it limits sensitivity at low
frequency. So, sensitivity below shot noise is achieved at high frequency
(adapted from Virgo-Ego Scientific Forum 2012 summer school lecture slides by
Stefan Hild, University of Glasgow). }%
\label{Squeeze}%
\end{figure}

In the real situation, the radiation pressure noise is significant only at low
frequencies (below 50 Hz or so) and the photon shot noise dominates the high
frequency region of the detection band. The physical picture is that the
vacuum noise enters the output port of the interferometer and adds to the
gravitational wave signal in the phase quadrature. Hence, any squeezing of the
phase quadrature, at the expense of increased noise in the amplitude
quadrature, reduces noise in the high frequency detection band where back
action from the amplitude quadrature through radiation pressure noise on the
mirror is insignificant (figure \ref{Squeeze}). \ This is then equivalent to
the use of higher laser power (more photons) in the interferometer, reducing
the quantum shot noise. However at low frequencies, the increased noise in the
amplitude quadrature will cause increased noise for gravitational wave
detection. This can be avoided only by frequency dependent squeezing, where
the phase quadrature is squeezed at high frequencies and amplitude quadrature
is squeezed at low frequencies. Implementation of sensitivity significantly
below shot noise in the relevant detection band is yet to be demonstrated in
full scale GW detectors, but feasibility has been demonstrated in these very
detectors at high frequency \cite{GEO-Squeeze,LIGO-Squeeze}.

\section{Renewed relevance of QND Measurements}

The efforts to detect gravitational waves have shifted focus from cryo-cooled
resonant detectors to interferometer based detectors with free mirrors as the
sensing masses. In such detectors, the expected displacement of the masses is
less than $10^{-19}$ m, which is smaller than the quantum zero-point motion of
these suspended mirrors. More seriously, the thermal motion is over a million
times larger, unlike in the cryo-cooled bar detectors where residual thermal
and quantum motions are comparable! However, effective metrology is possible
because the pendular suspensions of the mirrors have very high Q (quality
factor) and nearly the entire thermal and quantum energies are concentrated at
the oscillation frequency of about 1 Hz. Non-dissipative feed-back techniques
are used to keep these motions within certain limits and the actual detection
bandwidths starts 20-30 times higher in frequency \ where the residual from
the quantum and thermal motions are below the levels that can affect the
measurement. So, there is a clear separation between the detection bandwidth
and resonance bandwidth, in contrast to the resonant detectors where both
merge. Since resonant bar GW detector was the only metrological scenario that
necessarily needed a QND-BAE measurement for its success when these ideas
originated, one may wonder about the relevance of such ideas in the context of
advanced interferometer detectors. However, as we have seen, the
interferometric measurement is also limited by quantum noise in the optical
phase and amplitude quadratures and QND techniques with squeezed light is
turning out to be essential for the efficient operation of such detectors.
Also, QND metrology may significantly improve sensitivity and bandwidth in
magnetometry and rotation sensing (atomic gyroscopes) with spin-polarized
atomic ensembles. Another area of application where QNDM is indispensable is
in feed-back cooling of macroscopic oscillators to their quantum ground state
\cite{Hertzberg,Vanner}, which requires back-action evading measurements for
noise-free feedback.

\section{Summary remarks}

A survey of the experimental implementations of \ quantum non-demolition
measurements with back-action evasion, nearly four decades after such ideas
were first proposed, suggests that QNDM is maturely understood and have been
demonstrated in multiple physical systems. QNDM is demonstrated to be a useful
superior tool in those situations where metrology has done close to the
quantum noise level. Implementations are now a growing list including high
precision magnetometry and several types of optical measurements. QNDM is
crucially useful when not even measurements at the standard quantum limit can
take one to the goal of the measurement, as in the gravitational wave
detectors. Squeezed light technology as implemented in optical interferometers
may prove to be the single most important technology push that is required to
usher in gravitational wave astronomy.

\bigskip

\noindent\textbf{Acknowledgement: }The invitation from Prof. N. D. Haridass
and other organizers to the `Discussion Meeting on Quantum Measurements' was
an important opportunity for me to get more familiar with quantum
non-demolition measurements, which I believe will play a significant role in
future precision metrology below conventional quantum limit. However, I do not
think that this brief review does justice to the vast amount of published work
by several researchers, but I hope it will serve as a useful pointer.

\end{document}